**Title:**

A triangular triple quantum dot with tunable tunnel couplings

**Authors:**

A. Noiri[1,2], K. Kawasaki[1], T. Otsuka[2,3], T. Nakajima[2], J. Yoneda[2], S. Amaha[2], M. R. Delbecq[4], K. Takeda[2], G. Allison[2], A. Ludwig[5], A. D. Wieck[5], and S. Tarucha[1,2]

**Affiliations:**

[1]*Department of Applied Physics, University of Tokyo, 7-3-1 Hongo, Bunkyo-ku, Tokyo 113-8656, Japan*

[2]*RIKEN, Center for Emergent Matter Science (CEMS), Wako-shi, Saitama 351-0198, Japan*

[3]*JST, PRESTO, 4-1-8 Honcho, Kawaguchi, Saitama, 332-0012, Japan*

[4]*Quantum Matter Group, Collège de France, Paris, France*

[5]*Lehrstuhl für Angewandte Festkörperphysik, Ruhr-Universität Bochum, D-44780 Bochum, Germany*

**Abstract:**

A two-dimensional arrangement of quantum dots with finite inter-dot tunnel coupling provides a promising platform for studying complicated spin correlations as well as for constructing large-scale quantum computers. Here, we fabricate a tunnel-coupled triangular triple quantum dot with a novel gate geometry in which three dots are defined by positively biasing the surface gates. At the same time, the small area in the center of the triangle is depleted by negatively biasing the top gate placed above the surface gates. The size of the small center depleted area is estimated from the Aharonov-Bohm oscillation measured for the triangular channel but incorporating no gate-defined dots, with a value consistent with the design. With this approach, we can bring the neighboring gate-defined dots close enough to one another to maintain a finite inter-dot tunnel coupling. We finally confirm the presence of the inter-dot tunnel couplings in the triple quantum dot from the measurement of tunneling current through the dots in the stability diagram. We also show that the charge occupancy of each dot and that the inter-dot tunnel couplings are tunable with gate voltages.

**Introduction:**

Spins in semiconductor quantum dots (QDs) attract a lot of interest for their physical properties and applications to quantum computing hardware. The two-dimensional (2D) arrangement of tunnel-coupled QDs is intriguing for studies on complicated spin correlations including geometrical frustration of energetically degenerate states [1-5], underlying physics of high critical temperature cuprate superconductors [6] and applications to fault tolerant scalable quantum computing [7-10]. Triangles and squares with QDs at each apex are the basis for the 2D arrangement. Especially, a triangular-shaped triple quantum dot (TTQD) provides the simplest platform for studying the geometrical frustration since the frustration manifests itself even in a single unit cell [1-5]. In all these studies, inter-dot tunnel coupling is a key parameter. Gate-defined two QDs made in a 2D electron gas (2DEG) show a striking controllability of inter-dot tunnel couplings [11,12] but multiple tunnel-coupled QDs have been realized mostly in a 1D arrangement [13-15]. Two-dimensional arrangement of QDs was initially demonstrated using dry etching [16] and local anodic oxidation [17,18], but the inter-dot tunnel coupling is hard to vary in these devices since it is mainly determined by the device geometry and indeed exhibits no electrical tunability. More recently, a TTQD [5] and a square-shaped quadruple quantum dot [19] have been fabricated in a 2DEG using a surface electrical gating technique. However, it remains challenging to maintain the inter-dot tunnel couplings mainly because QDs made by negative surface gating tend to be largely spaced from each other (~ 500 nm). In this work, we take a different approach to realize TTQDs with a distance between QDs short enough (~ 250 nm) to maintain the finite and gate-tunable inter-dot tunnel coupling. The main improvement of our device is the realization of the smaller depletion region in the center of TTQDs which enables us to reduce the inter-dot distance. To this end we employ two key techniques. Firstly, we apply positive voltages to three surface plunger gates to form three closely spaced QDs underneath. Secondly, to deplete the center of the TTQD device, we negatively bias a top gate placed above the surface gates. We evaluate the size of the center depleted area by calculating the electrostatic potential of the TTQD device and from measurements of the AB oscillations for the device incorporating no gate-defined quantum dots, and finally measure tunneling current through all three dots defined by gating to confirm the electrical tuning of the inter-dot tunnel couplings. The approach presented in this work may be scaled up to form 2D QD arrays.

**Results:**

Our device is a gate-defined lateral TTQD fabricated in a 2DEG formed at a GaAs/AlGaAs hetero-interface 100 nm below the surface (Fig. 1(a)). We apply negative voltages to the surface gates to define the outer boundaries of the three dots, and positive voltages to the three plunger gates (P1, P2, P3) to place the three dots close to each other. We label the three dots as QD1, QD2, and QD3 formed under gates P1, P2, and P3, respectively. The inter-dot separation is designed to be 250 nm, which is determined by the distance between the centers of the neighboring plunger gates. This distance is short enough to have a finite tunnel coupling for the 2DEG used based on previous work [20,21]. The top gate is placed over the TTQD structure with a 50nm insulator in between and, when negatively biased, it depletes the center area surrounded by the three plunger gates. Note the plunger gates screen the electric field imposed by the top gate, and therefore we can deplete predominantly the center area without undesirable spreading of the depletion region outward. This approach may be applicable to 2D arrays of Si QDs to have a large impact on large-scaling of quantum circuits. All of the surface gates and the top gate are designed based on the numerical calculation of the electrostatic potential as shown in Fig. 1(b) [22]. We see three potential valleys corresponding to the three QDs by applying positive voltages to the plunger gates and negative voltages to the other gates. All measurements are performed using a dilution fridge at a base temperature around 30 mK and with an out-of-plane magnetic field $B$ ranging from 0 to 3.5 T.

We first examine the depletion of the center area while keeping the three dots strongly connected to each other making two current channels. We measure the current $I$ flowing between the two Ohmic contacts O3 and O6 under a source-drain bias of $V_{SD} = 200$ μV. The two-terminal conductance $G$ arises from the two current paths: the upper channel of O3 → QD1 → QD2 → O6 and the lower channel of O3 → QD1 → QD3 → QD2 → O6 as shown in the inset of Fig. 2(a). Here, we set the QD-lead and inter-dot couplings to be large such that three QDs are no longer formed but two current paths are well established. Fig. 2(a) shows $G$ measured as a function of gate voltage $V_G$ applied commonly to the three gates, D1, D2, and T3. While $G$ decreases as $V_G$ is decreased from -0.5 V, the change of $G$ is found to be rather abrupt in -0.7 V > $V_G$ > -0.8 V and $V_G$ < -1.2 V. The first abrupt drop of $G$ is attributed to the depletion of the lower current channel and the second to that of the upper current channel. The gradual decrease of $G$ in between is due to the screening by the plunger gates P1 and P3. Therefore, the transport occurs through the two channels in the region of -0.5 V > $V_G$ > -0.7 V (highlighted in red in Fig. 2 (a)) and only through the upper channel in the region of -0.8 V > $V_G$ > -1.2 V (highlighted in blue). The two channels have more or less the same conductance judging from the $G$ values found in the one- and two-channel transport.

Then we measure $B$ dependence of $G$ in two different $V_G$ regions for the two-channel transport and the one channel transport (see the red and blue regions of Fig. 2(a), respectively). The result is plotted in Fig. 2(b). $G$ decreases at high $B$ due to the quantum Hall effect in the contact regions [23]. This is predicted to appear for $B > 1.9$ T. In the lower $B$ field range, we observe an oscillatory behavior particularly for the two-channel transport with $V_G$ = -0.5 and -0.6 V. The sharp dip at $B = 0$ T is attributed to weak localization [24]. Neglecting this sharp dip, relatively broad peaks are located at $B$ = 0, ±1.5 T. We attribute these peaks to Aharonov-Bohm (AB) oscillations whose period is determined by the flux quantum penetrating the depleted area in the center [25]. Naively, we expect that the depleted area corresponds to the region underneath the top gate but uncovered by the plunger gate metals. From the obtained oscillation period of ~ 1.5 T we evaluate the central depleted area of an 80 nm equilateral triangle (yellow dashed triangle in Fig. 1(a)), which roughly compares with the area we expect. Again, note that the $B = 0$ T dip is too narrow to be attributed to a part of the AB oscillations. For the one-channel transport with $V_G$ = -1.1 and -1.2 V the peak feature is mostly obscured, also as expected.

Then we tune all gate voltages appropriately to form a TTQD contacted to two leads, O3 and O9 as shown in Fig. 3(a). Here no magnetic field is applied. We measure transport current flowing between the two contacts through the TTQD as a function of P1 and P2 gate voltages to obtain the stability diagram (Fig. 3(b)). The measured current is the sum of the current flowing through the two paths: O3 → QD1 → QD3 → O9 for Path1 ($I_1$) and O3 → QD1 → QD2 → QD3 → O9 for Path2 ($I_2$) as shown in Fig. 3(a). Formation of three dots is confirmed by observing three families of the charging lines with different slopes: large, intermediate, and small for the charging of QD1, QD2, and QD3, respectively. Here, the QD-lead and inter-dot tunnel couplings are relatively strong so that the charge state boundaries are observed as lines in the transport current due to co-tunneling, not isolated spots on the triple points. The observation of the three families of charging lines for three QDs by the transport measurement indicates the existence of finite inter-dot tunnel couplings between all pairs of the QDs.

To study the tunability of the inter-dot tunnel couplings, we evaluate $I_1$ and $I_2$ as a function of T1, T2, and T3 gate voltages. Along the Coulomb peak of QD1 in Fig. 3(b), we see the current oscillation as shown in Fig. 4(a). This is mainly due to the Coulomb oscillation for charging QD2 one-by-one. We assign the current value at a valley to $I_1$ where the transport through Path2 is prohibited by the Coulomb blockade of QD2. On the other hand, $I_2$ can be extracted by subtracting $I_1$ from the peak current as indicated by dashed lines. The ratio of the two values, $I_{2/1} = I_2 / I_1$, extracted from the peak indicated by the black circle in Fig. 4(a) is shown in Figs. 4 (b), (c), and (d) as a function of T1, T2, and T3 gate voltage, respectively. The values of $I_1$ and $I_2$ are also plotted. T1, T2, and T3 are assumed to predominantly affect the inter-dot tunnel couplings of QD1-2, QD2-3, and QD3-1, respectively, due to the screening effect of P1, P2 and P3 gate metals. Figs. 4 (b) and (c) show that $I_2$ gradually decreases with decreasing T1 and T2 voltages, whereas $I_1$ is only weakly changed. Accordingly, $I_{2/1}$ apparently decreases with decreasing voltage on T1 (T2) in (b) ((c)). This is because $I_2$ is more efficiently suppressed in deeper T1 and T2 gate voltages than $I_1$. Therefore, $I_2$ is significantly affected by the T1 and T2 gates, while $I_1$ should be strongly affected by the T3 gate. As expected, $I_1$ decreases more rapidly and $I_{2/1}$ increases with decreasing T3 gate voltage as shown in Fig. 4 (d) compared to that in (b) and (c). From all these observations we confirm that the inter-dot tunnel couplings between each QD are tunable with gate voltages. Note the data sets in Fig. 4 (b) to (d) are extracted from a specific peak in Fig. 4(a) but the trends observed are more or less the same for the other peaks in Fig. 4(a). This is the first observation of a triangular triple QD with gate tunable inter-dot tunnel couplings, and our results indicate that the gate geometry employed here is useful to prepare the 2D arrangement of multiple QDs with keeping tunable inter-dot tunnel couplings. Though the few electron regime is not reached in this experiment, it may be achievable by reducing the size of the dot design.

Although we did not measure a magnetic field dependence with QDs in this study, we anticipate the AB effect can be observed even when QDs are formed in the two paths of Fig. 3(a). AB oscillations may offer an interesting subject to study the coherent transport through one or two QDs such as a phase shift through QDs [26] and a phase difference between two electron states of singlet and triplet [27,28].

**Conclusion:**

In conclusion, we realize a gate-defined TTQD device in a GaAs 2DEG with electrically tunable inter-dot tunnel couplings. We apply positive voltages to three plunger gates to form three closely separated QDs underneath while depleting the small center area surrounded by the three dots by negatively biasing a gate on top. The size of the center depletion area, which is estimated from the AB oscillation measured for the device with no gate-defined dots is as small as predicted by the design, and therefore we can sufficiently reduce the distance between gate-defined QDs to maintain the finite inter-dot tunnel couplings. We measure tunneling current through the three dots to establish a stability diagram and demonstrate that the inter-dot tunnel couplings between all pairs of QDs and electron occupation of each dot are both tunable with gate voltages. The gate geometry employed in this work may thus be useful for making 2D QD arrays to study complicated electron correlations and applications to scalable QD-based architecture for quantum computing.

**Acknowledgement:**

Part of this work is financially supported by the ImPACT Program of Council for Science, Technology and Innovation (Cabinet Office, Government of Japan) the Grant-in-Aid for Scientific Research (No. 26220710, 16H00817), CREST (JPMJCR15N2, JPMJCR1675), PRESTO (JPMJPR16N3), JST, Incentive Research Project from RIKEN. AN acknowledges support from Advanced Leading Graduate Course for Photon Science (ALPS). TN acknowledges financial support from JSPS KAKENHI Grant Number 25790006. TO acknowledges financial support from JSPS KAKENHI Grant Number 25800173, Strategic Information and Communications R&D Promotion Programme, Yazaki Memorial Foundation for Science and Technology Research Grant, Japan Prize Foundation Research Grant, Advanced Technology Institute Research Grant, the Murata Science Foundation Research Grant, Izumi Science and Technology Foundation Research Grant, TEPCO Memorial Foundation Research Grant, The Thermal & Electric Energy Technology Foundation Research Grant, The Telecommunications Advancement Foundation Research Grant. AL and ADW acknowledge gratefully support of DFG-TRR160, BMBF - Q.com-H 16KIS0109, and the DFH/UFA CDFA-05-06.

**Figures:**

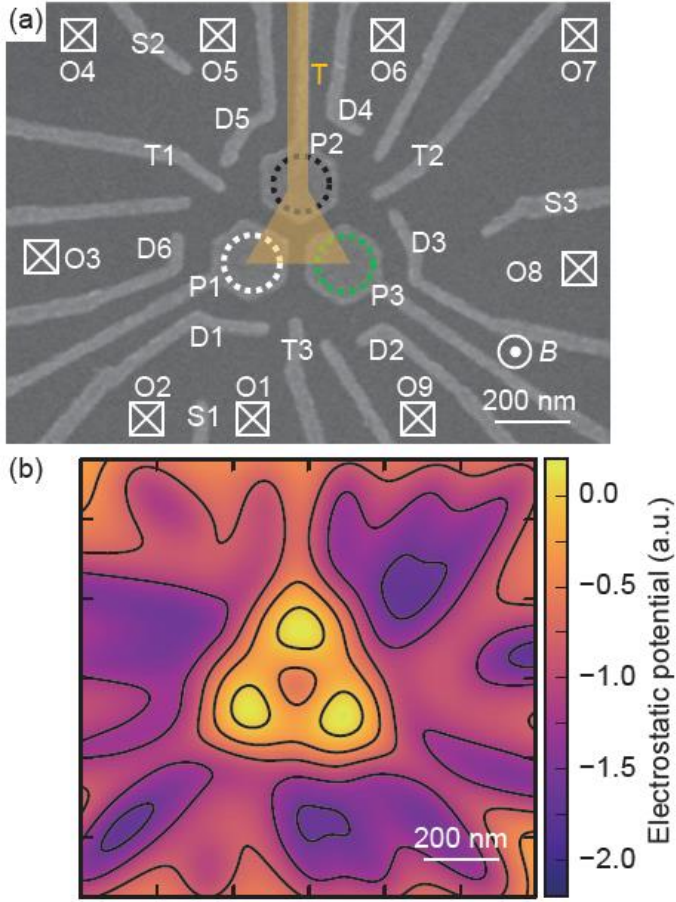

Fig. 1 (a) Scanning electron micrograph (SEM) of a TTQD device prepared in the same way as the one used in the experiment. QD1, QD2, and QD3 are formed under the three plunger gates P1, P2, and P3, respectively. Another gate in orange is placed on top to deplete the center area. The yellow dashed triangle shows the estimated depletion area (explained in the text). O3, O6, and O9 are Ohmic contacts to leads used as a lead for QD1, QD2, and QD3, respectively. Gates T1, T2, and T3 change the inter-dot tunnel couplings. Gates D1 to D6 define the outer boundaries of the three dots. Other contacts and gates S1, S2, and S3 are designed to perform charge sensing, but they are not used in this work. (b) Numerical calculation of the electrostatic potential created by the surface gates and the top gate [22] in (a). In this calculation, no screening effect by the 2DEG is included. We assume that +0.4 V is applied for the three plunger gates and -1.0 V for the other surface gates and the top gate for the calculation. Yellow and purple regions show the potential valleys, and peaks, respectively.

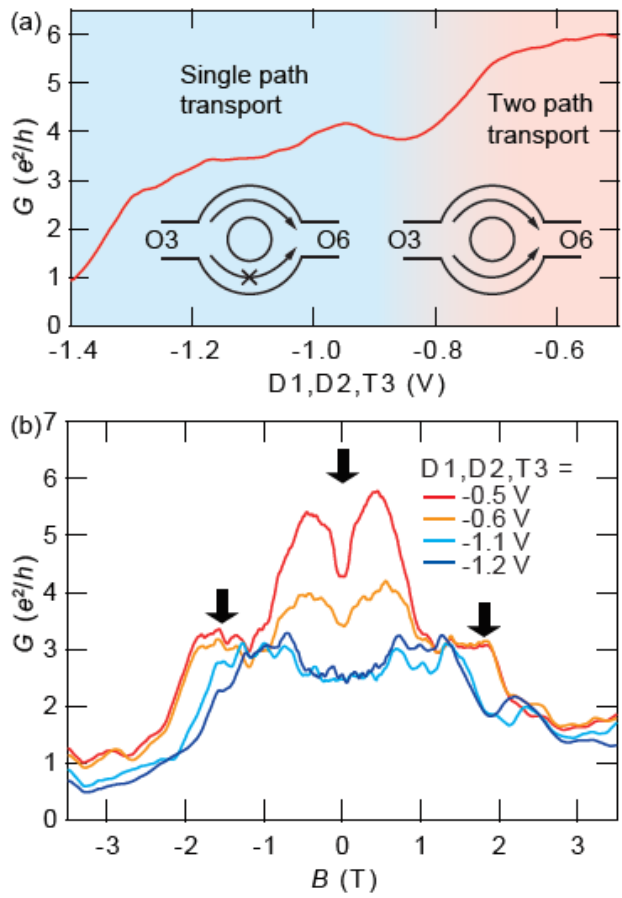

Fig. 2 (a) Conductance $G$ between O3 and O6 as a function of voltage $V_G$ applied to gates D1, D2, and T3. The red (blue) region shows the regime of the two (single) path transport as indicated in the right (left) lower inset. (b) $G$ measured as a function of out-of-plane magnetic field $B$ with $V_G$ = -0.5, -0.6, -1.1, and -1.2 V. The black arrows show the conductance peak positions expected for the AB oscillation (see text).

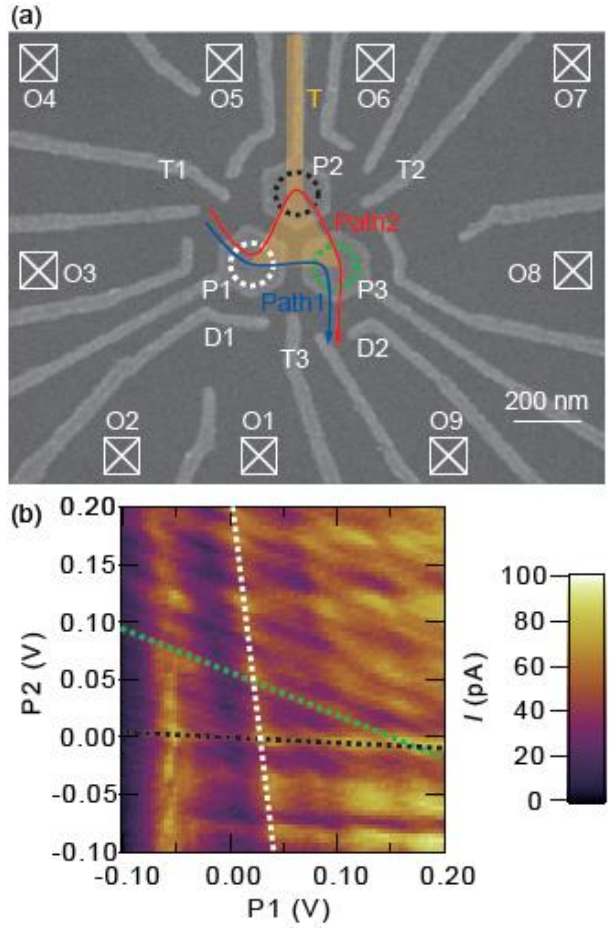

Fig. 3 (a) Schematic of the TTQD transport paths in the measurement setup. The blue and red lines indicate the paths of QD1-3 (Path1), and QD1-2-3 (Path2). (b) TTQD stability diagram taken as a function of P1 and P2 gate voltages by transport measurement between O3 and O9 with $V_{SD} = 50$ μV. White, black, and green dashed lines correspond to the single electron charging of QD1, QD2, and QD3, respectively. The number of electrons in each QD is not determined because it is difficult to trace the charge states down to the few-electron regime by transport measurement.

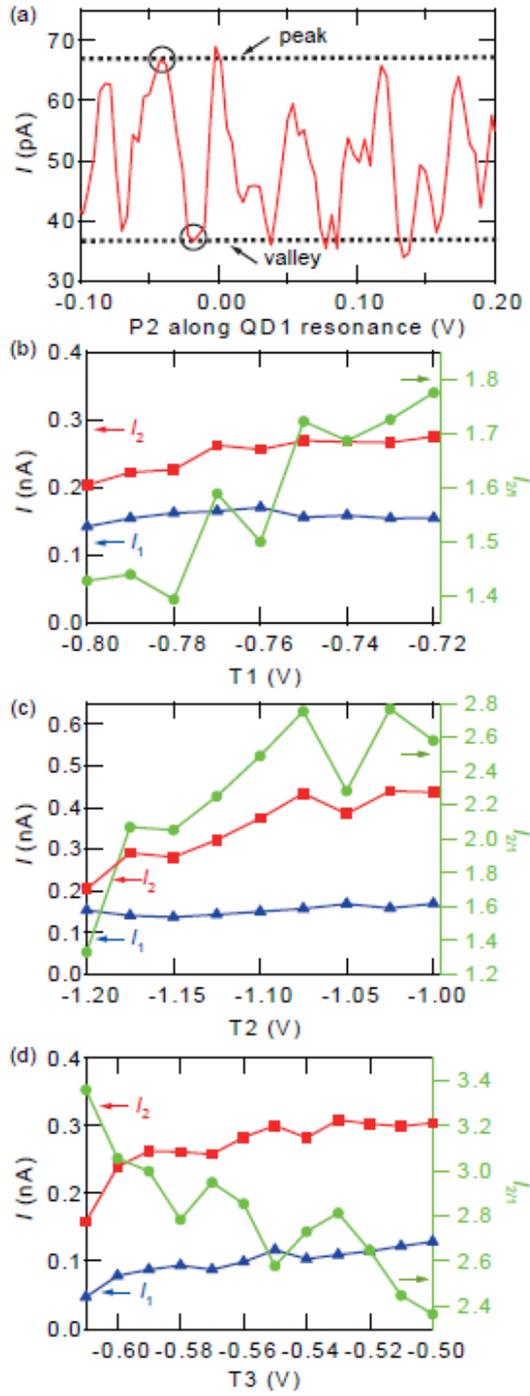

Fig. 4 (a) Transport current along the Coulomb peak of QD1 as indicated by the white dashed line in Fig. 3(b). The Coulomb peak and valley, and their current levels used for the analysis are indicated by black circles, and dashed lines, respectively. (b) – (d) Gate voltage dependence of each contribution from the two paths is shown for T1 in (b), T2 in (c), and T3 in (d), respectively. The left axis shows the current through Path1 ($I_1$) and Path2 ($I_2$), respectively. The right axis shows the ratio of the two, $I_{2/1} = I_2 / I_1$.

13